\documentclass[conference]{IEEEtran}
\usepackage{xurl}
\usepackage[noadjust]{cite}
\usepackage{graphicx}
\usepackage{algpseudocode}
\usepackage{algorithm}
\usepackage{tabularx}
\usepackage{makecell}
\usepackage[table]{xcolor}
\newcolumntype{P}[1]{>{\centering\arraybackslash}p{#1}}
\newcolumntype{M}[1]{>{\centering\arraybackslash}m{#1}}
\usepackage[edges]{forest}
\usepackage{tikz}
\usepackage{pgfplots}
\pgfplotsset{compat=1.17}
\pgfdeclarelayer{background}
\pgfsetlayers{main}
\usepackage{multirow}
\usetikzlibrary{external}
\usepackage{rotating}
\usepackage{amsmath}

\usepackage{algorithm, algpseudocode}
\usepackage{setspace}
\usepackage{xcolor}
\usepackage{gensymb}
\usepackage{placeins}
\usepackage{enumitem}
\usetikzlibrary{decorations.pathreplacing,positioning}
\usepackage[noadjust]{cite}
\usepackage{dirtytalk}

\definecolor{dkgreen}{rgb}{0,0.6,0}
\definecolor{gray}{rgb}{0.5,0.5,0.5}
\definecolor{mauve}{rgb}{0.58,0,0.82}
\definecolor{darkblue}{rgb}{0.0,0.0,0.6}
\definecolor{cyan}{rgb}{0.0,0.6,0.6}
\definecolor{mBlue}{HTML}{4285f4}
\definecolor{mRed}{HTML}{ea4335}
\definecolor{mGreen}{HTML}{34a853}
\definecolor{mYellow}{HTML}{fbbc04}
\definecolor{mLightBlue}{HTML}{6d9eeb}
\definecolor{mLightRed}{HTML}{e06666}
\definecolor{mLightGreen}{HTML}{93c47d}
\definecolor{mLightYellow}{HTML}{ffd966}
\definecolor{archtBlue}{HTML}{5DADE2}
\definecolor{archtTeal}{HTML}{a4d2d7}
\definecolor{archtYellow}{HTML}{ffe599}
\definecolor{archtOrange}{HTML}{E59866}
\definecolor{archtCetus}{HTML}{ea9999}
\definecolor{archtOther}{HTML}{dd7e6b}
\definecolor{archtPurple}{HTML}{b4a7d6}
\definecolor{archtGrey}{HTML}{eeeeee}
\definecolor{archtGreen}{HTML}{b6d7a8}
\definecolor{archtRed}{HTML}{ea9999}

\usepackage{tikz}
\usepackage{rotating}
 \usepackage[frozencache,cachedir=mint]{minted}
\usepackage{hyperref}
\usepackage{pgf-pie}  
\usepackage{listings}
\usepackage{media9}
\usepackage{lipsum}

\usepackage{etoolbox}

\newcommand{\hpcorpus}[0]{\textsc{HPCorpus}}

\definecolor{codegreen}{rgb}{0,0.6,0}
\definecolor{codegray}{rgb}{0.5,0.5,0.5}
\definecolor{codepurple}{rgb}{0.58,0,0.82}
\definecolor{backcolour}{rgb}{0.95,0.95,0.92}

\lstdefinestyle{mystyle}{
    backgroundcolor=\color{backcolour},   
    commentstyle=\color{codegreen},
    keywordstyle=\color{magenta},
    numberstyle=\tiny\color{codegray},
    stringstyle=\color{codepurple},
    basicstyle=\ttfamily\footnotesize,
    breakatwhitespace=false,         
    breaklines=true,                 
    captionpos=b,                    
    keepspaces=true,                 
    numbers=left,                    
    numbersep=5pt,                  
    showspaces=false,                
    showstringspaces=false,
    showtabs=false,                  
    tabsize=2
}
\usepackage{subfigure}
\usepackage{tcolorbox}
\usepackage{pdfpages}

\usetikzlibrary{shapes, arrows}
\usetikzlibrary{matrix, positioning, fit}

\DeclareRobustCommand*{\IEEEauthorrefmark}[1]{%
  \raisebox{0pt}[0pt][0pt]{\textsuperscript{\footnotesize\ensuremath{#1}}}}
  \usepackage[misc]{ifsym}

\ifCLASSINFOpdf

\else

\fi

\DeclareRobustCommand*{\IEEEauthorrefmark}[1]{%
  \raisebox{0pt}[0pt][0pt]{\textsuperscript{\footnotesize\ensuremath{#1}}}}
  \usepackage[misc]{ifsym}

\begin{document}
\title{Quantifying OpenMP:\\Statistical Insights into Usage and Adoption}
\author{\IEEEauthorblockN{Tal Kadosh\IEEEauthorrefmark{1,2},
Niranjan Hasabnis\IEEEauthorrefmark{3},
Timothy Mattson\IEEEauthorrefmark{3},
Yuval Pinter\IEEEauthorrefmark{1} and
Gal Oren\IEEEauthorrefmark{4,5}}\\
\IEEEauthorblockA{\IEEEauthorrefmark{1}Department of Computer Science, Ben-Gurion University, Israel}
\IEEEauthorblockA{\IEEEauthorrefmark{2}Israel Atomic Energy Commission}
\IEEEauthorblockA{\IEEEauthorrefmark{3}Intel Labs, United States}
\IEEEauthorblockA{\IEEEauthorrefmark{4}Scientific Computing Center, Nuclear Research Center – Negev, Israel}
\IEEEauthorblockA{\IEEEauthorrefmark{5}Department of Computer Science, Technion – Israel Institute of Technology, Israel}
{\tt\small talkad@post.bgu.ac.il, niranjan.hasabnis@intel.com,}\\ {\tt\small timothy.g.mattson@intel.com, pintery@bgu.ac.il, galoren@cs.technion.ac.il}
}
\IEEEtitleabstractindextext{%
\begin{abstract}
In high-performance computing (HPC), the demand for efficient parallel programming models has grown dramatically since the end of Dennard Scaling and the subsequent move to multi-core CPUs. OpenMP stands out as a popular choice due to its simplicity and portability, offering a directive-driven approach for shared-memory parallel programming. Despite its wide adoption, however, there is a lack of comprehensive data on the actual usage of OpenMP constructs, hindering unbiased insights into its popularity and evolution.

This paper presents a statistical analysis of OpenMP usage and adoption trends based on a novel and extensive database, \hpcorpus{}, compiled from GitHub repositories containing C, C++, and Fortran code. The results reveal that OpenMP is the dominant parallel programming model, accounting for 45\% of all analyzed parallel APIs. Furthermore, it has demonstrated steady and continuous growth in popularity over the past decade. Analyzing specific OpenMP constructs, the study provides in-depth insights into their usage patterns and preferences across the three languages. Notably, we found that while OpenMP has a strong ``common core'' of constructs in common usage (while the rest of the API is less used), there are new adoption trends as well, such as \texttt{simd} and \texttt{target} directives for accelerated computing and \texttt{task} for irregular parallelism. 

Overall, this study sheds light on OpenMP's significance in HPC applications and provides valuable data for researchers and practitioners. It showcases OpenMP's versatility, evolving adoption, and relevance in contemporary parallel programming, underlining its continued role in HPC applications and beyond. These statistical insights are essential for making informed decisions about parallelization strategies and provide a foundation for further advancements in parallel programming models and techniques.

\hpcorpus{}, as well as the analysis scripts and raw results, are available at: 
\url{https://github.com/Scientific-Computing-Lab-NRCN/HPCorpus}
\\
\end{abstract}
}
\maketitle
\IEEEdisplaynontitleabstractindextext
\IEEEpeerreviewmaketitle
\section{Introduction}

With the end of Dennard Scaling~\cite{esmaeilzadeh2011dark}, multicore CPUs sharing a cache-coherent address space are ubiquitous. To exploit the parallelism available from multicore systems, programmers use multithreaded programming models. Programming models that support multithreaded
execution include pThreads~\cite{buttlar1996pthreads} for low-level and OS-level operations, TBB~\cite{kukanov2007foundations} or Cilk~\cite{blumofe1995cilk} for task-level parallelism
in C++, and OpenMP~\cite{mattson1999openmp} for directive-driven parallelism.

Despite the popularity of multithreaded models, little empirical data is available to assess the actual usage of the various programming constructs from these models.
While anecdotal data and feedback from user-support teams at supercomputing centers exist~\cite{commonstats}, a large-scale analysis has yet to be conducted.
In this paper, we perform a statistical analysis of repositories from GitHub to study the usage of parallel programming models. Our analysis reveals that OpenMP is the dominant programming model for writing multithreaded applications. We also go inside applications to gather usage data on specific OpenMP constructs. Finally, we consider the evolution of OpenMP and the adoption of newly included constructs as OpenMP Specifications are released.
 

\section{OpenMP Fundamentals vs. Other Parallel Programming APIs}

In this section, we provide a brief overview of OpenMP and other parallel programming APIs.
OpenMP~\cite{mattson2019openmp} defines a simple and portable approach to shared-memory parallel programming.
OpenMP makes parallel programming more accessible by offering a directive-based approach, where directives inserted into the code guide the compiler as it  generates parallel code.
These directives provide high-level abstractions to specify parts of a code to execute in parallel.
OpenMP uses the fork-join model of parallelism, where a single thread (the \textsc{primary} thread) on encountering a parallel directive forks a team of threads, each of which executes the code in a parallel region independently.
Synchronization constructs, such as barriers, coordinate multithreaded execution, while shared variables facilitate data sharing among threads.
OpenMP emphasizes portability with a standardized API that can be used across different platforms, hardware, and programming languages. These features make it easier to write parallel code that can be compiled and executed on systems supporting OpenMP.

OpenMP differs from other parallel APIs in several ways.
Unlike Cilk and TBB, which focus on task-based parallelism, OpenMP is more general and addresses loop-level parallelism, task-parallelism, and general, multi-threaded parallelism through explicit thread-level programming. 
OpenMP differs from MPI~\cite{gabriel2004open}, which is designed for distributed-memory parallelism by targeting shared-memory parallelism within a single program.
Compared to CUDA~\cite{luebke2008cuda} and OpenCL~\cite{munshi2011opencl}, which are geared towards throughput-optimized accelerators (and in the case of CUDA, a specific vendor), OpenMP provides a portable solution that works across different platforms, including general-purpose CPUs, accelerators~\cite{schmidl2013assessing}, 
GPUs~\cite{van2017using,deakin2023programming, fridman2023portability}, and FPGAs~\cite{mayer2019openmp}. Additionally, OpenMP's compatibility with multiple programming languages (specifically C, C++, Fortran, and partially Python~\cite{anderson2021multithreaded,mattson2021pyomp}) sets it apart from other GPU programming languages such as SYCL~\cite{reyes2016sycl} (C++ only) and OpenCL (C and C++).
OpenACC~\cite{farber2016parallel} is similar to OpenMP in its directive-based approach, but it specifically targets GPUs only.  OpenACC emphasizes descriptive semantics, meaning constructs describe \textit{what} should be accomplished, not \textit{how} it is done.  OpenMP, on the other hand, emphasizes \textit{prescriptive} semantics, which allows the programmer to control how code maps onto a system explicitly.  With the \texttt{loop}-construct added in OpenMP version 5.0, however, OpenMP is moving to include descriptive semantics, thereby supporting OpenACC's approach for programmers who prefer yielding more control to the compiler.


\section{\hpcorpus{}: A Novel Database of HPC Code from GitHub}
\begin{figure}
    \centering

\begin{minted}[fontsize=\footnotesize, linenos, frame=single, firstnumber=1, breaklines]{sql}
WITH selected_repos as (
SELECT f.id, f.repo_name as repo_name, f.ref as ref, f.path as path
FROM `bigquery-public-data.github_repos .files` as f
JOIN `bigquery-public-data.github_repos .licenses` as l on l.repo_name = f.repo_name
),
deduped_files as (
SELECT f.id, MIN(f.repo_name) as repo_name, MIN(f.ref) as ref, MIN(f.path) as path
FROM selected_repos as f
GROUP BY f.id
)
SELECT
f.repo_name, f.ref, f.path, c.copies, c.content,
FROM deduped_files as f
JOIN `bigquery-public-data.github_repos .contents` as c on f.id = c.id
WHERE
 NOT c.binary
 AND (f.path like '%.c' OR f.path like '%.cpp' OR f.path like '%.f' OR f.path like '%.f90' OR f.path like '%.f95')
\end{minted}
    \caption{\hpcorpus{} data acquisition from Google's BigQuery.}
    \label{fig:hpcorpus}
\end{figure}




\begin{table}
\vspace{4pt}

    \centering
    \resizebox{0.5\textwidth}{!}{%
      \begin{tabular}{|c |c |c |c |c |} 
     \hline
     & \textbf{Repos} (\#)  & \textbf{Size} (GB) &  \textbf{Files} (\#) & \textbf{Functions} (\#)\\ 
     \hline
     \textbf{C} & 144,522  & 46.23 & 4,552,736 & 87,817,591\\
     \hline
     \textbf{C++} & 150,481 & 26.16 & 4,735,196 & 68,233,984\\
     \hline
     \textbf{Fortran} & 3,683 & 0.68 & 138,552 & 359,272\\
     \hline
\end{tabular}
}
\vspace{2pt}
    \caption{Total number of repositories in \hpcorpus{} by language. \\Note: repositories may use multiple languages, and 5.5K repositories in HPCorpus contained no code.}
    \label{table:hpcorpus_table}
\end{table}

To study the usage of parallel programming APIs, we compiled a novel  database called \hpcorpus{}. It collects C, C++, and Fortran codes from every publicly visible GitHub repository that was accessible via BigQuery with the suffixes c, cpp, f, f90, and f95 (\autoref{fig:hpcorpus}, line 17). These languages are widely recognized as the dominant languages in HPC~\cite{comp_1, comp_2, comp_3, comp_4, comp_5, comp_6, comp_7, comp_8, comp_9}.
In the data acquisition process, we followed~\cite{lachaux2020unsupervised, transcoder}  but selected the C, C++, and Fortran files within those projects in an unrestricted way.
We structured \hpcorpus{} with a JSON format using the script presented in \autoref{fig:hpcorpus}.\footnote{Although the script does not contain code to obtain repository timestamps, we did collect them separately using perceval~\cite{duenas2018perceval}. We store these timestamps separately to keep \hpcorpus{} to a more reasonable size.}

A breakdown of the repository 
statistics is presented in \autoref{table:hpcorpus_table}.
Fortran has far less data than the C and C++ sub-corpora. The C sub-corpus is almost twice as large as the C++ sub-corpus. 

\hpcorpus{} aims to achieve two main objectives: first, to facilitate statistical analyses on the popularity and usage of various parallel programming APIs, and second, to serve as a robust training resource for the next generation of Large Language Models (LLMs)~\cite{xu2022systematic} designed to automate complex high-performance tasks~\cite{chen2023lm4hpc}.
Of these tasks, parallelization is most notable~\cite{harel2023learning, kadosh2023advising, schneider2023mpi, godoy2023evaluation, nichols2023modeling}, since current rule-based compilers for such purposes are not optimal or robust~\cite{harel2020source, mosseri2020compar}.
By offering a diverse and vast array of code snippets, \hpcorpus{} has the potential to significantly enhance research and development efforts in advanced AI models that address intricate parallelization challenges.


\section{Analysis}

We divide our analysis into two parts. First, we compare the popularity of OpenMP as a function of total usage and usage over time relative to  other parallel programming APIs.
Next, we present specific statistics about OpenMP constructs and the extent to which users adopt them.


\subsection{OpenMP vs. Other Parallel Programming APIs}

In this section, we elaborate on the usage statistics of OpenMP vs. other common parallel programming APIs, such as MPI, CUDA, OpenCL, TBB, Cilk, OpenACC, and SYCL (\autoref{fig:merged_figure_hpcorpus}\footnote{See code @ \href{https://github.com/Scientific-Computing-Lab-NRCN/HPCorpus/blob/121b0f088f3187cf82a54016e82992400cfcfe2c/generate_stats.py\#L100}{aggregate\_paradigms}.}). 
We find that not only is OpenMP by far the most popular API, but it accounts for almost half (45\%) of all the repositories using parallel programming APIs in this analysis.
Over the last decade (since 2013), there has been steady growth in new OpenMP repositories.
When combined with absolute usage numbers for OpenMP, it maintains a clear popularity advantage relative to other APIs (\autoref{fig:repo_per_year}).\footnote{See code @ \href{https://github.com/Scientific-Computing-Lab-NRCN/HPCorpus/blob/121b0f088f3187cf82a54016e82992400cfcfe2c/generate_stats.py\#L179}{get\_paradigms\_per\_year}.}
We also observe that MPI, as a distributed-memory parallelization API, and OpenMP, as a shared-memory parallelization API, are well integrated over time (\autoref{fig:omp-mpi}),\footnote{See code @ \href{https://github.com/Scientific-Computing-Lab-NRCN/HPCorpus/blob/121b0f088f3187cf82a54016e82992400cfcfe2c/generate_stats.py\#L193}{get\_omp\_mpi\_usage}.}
indicating the growing need for MPI+X as parallel clusters with multicore nodes grow in scale~\cite{thakur2010mpi}.


\begin{figure}[!htbp]
\centering
\begin{tikzpicture}
\begin{axis}  
[  
    ybar,
    ymax=4500,
    enlargelimits=0.15, 
    ylabel={\# Repos},
    symbolic x coords={
        OpenMP,
        MPI,
        OpenCL,
        CUDA,
        TBB,
        Cilk,
        OpenACC,
        SYCL
    },
    xtick=data,
    nodes near coords,
    every node near coord/.append style={rotate=90, anchor=west},
    xticklabel style={rotate=40}
]  
\addplot[fill=archtGrey] coordinates {
    (OpenMP, 3881)
    (MPI, 2340)
    (CUDA, 879)
    (OpenCL, 956)
    (OpenACC, 61)
    (SYCL, 24)
    (TBB, 374)
    (Cilk, 97)
}; 
\end{axis}

\begin{scope}[shift={(4.3cm,3.7cm)},scale=0.6]
\pie[color={archtOrange,archtBlue,archtRed,archtTeal,archtPurple}]{
    45/OpenMP,
    27.1/MPI,
    11.1/OpenCL,
    10.2/CUDA,
    6.6/Others
}
\end{scope}

\end{tikzpicture}  
\caption{Parallel programming API usage in \hpcorpus{}.}
\label{fig:merged_figure_hpcorpus}
\end{figure}
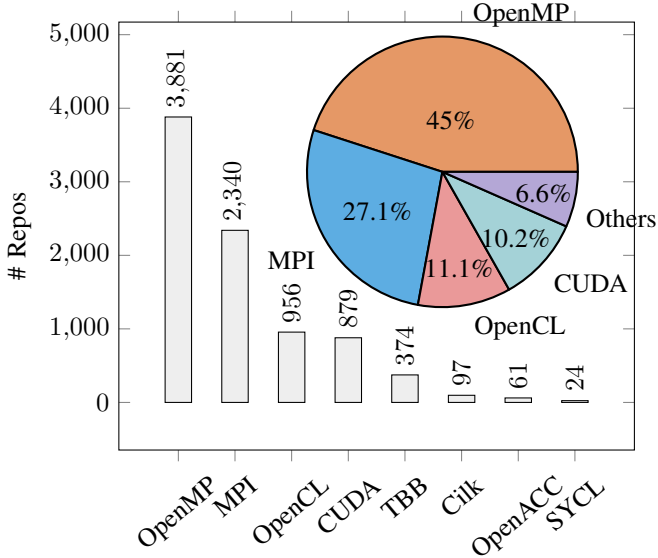

\begin{figure}[t!]
    \centering
    \begin{tikzpicture}
    \begin{axis}[
        legend columns=4,
        width=0.45\textwidth,
        height=0.3\textwidth,
        xlabel={Year},
        ylabel={\# Repos (log scale)},
        xmin=2012, xmax=2024,
         ymax=1200,
        xtick={2013,2016,...,2024},
        ymode=log,
        log basis y=10,
        legend pos=north west,
        ymajorgrids=true,
        xmajorgrids=true,
        xticklabel style={/pgf/number format/1000 sep=\,},
        xlabel style={yshift=-5pt},
        legend style={at={(0.5,-0.3)},anchor=north,legend columns=-1},
        grid style=dashed,
    ]

    \addplot[color=archtOrange,mark=diamond,thick] table[x=Year,y=OpenMP,col sep=comma] {results/repos_per_year.update.csv};
    \addplot[color=archtBlue,mark=diamond,thick] table[x=Year,y=MPI,col sep=comma] {results/repos_per_year.update.csv};
    \addplot[color=archtRed,mark=diamond,thick] table[x=Year,y=OpenCL,col sep=comma] {results/repos_per_year.update.csv};
    \addplot[color=archtTeal,mark=diamond,thick] table[x=Year,y=CUDA,col sep=comma] {results/repos_per_year.update.csv};
    \addplot[color=archtPurple,mark=diamond,thick] table[x=Year,y=TBB,col sep=comma] {results/repos_per_year.update.csv};
    \addplot[color=black,mark=diamond,thick] table[x=Year,y=Cilk,col sep=comma] {results/repos_per_year.update.csv};
    \addplot[color=mLightGreen,mark=diamond,thick] table[x=Year,y=OpenACC,col sep=comma] {results/repos_per_year.update.csv};
    \addplot[color=mGreen,mark=diamond,thick] table[x=Year,y=SYCL,col sep=comma] {results/repos_per_year.update.csv};
    
    \legend{OpenMP,MPI,OpenCL,CUDA,TBB,Cilk,OpenACC,SYCL} 
    \end{axis}
    \end{tikzpicture}  
    \caption{Parallel programming API usage trends in \hpcorpus{} over the last decade. }
    \label{fig:repo_per_year}
\end{figure}
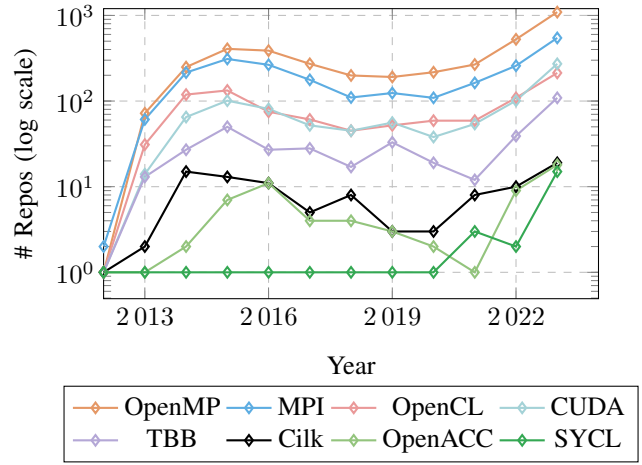

\begin{figure*}
    \centering
    \begin{tikzpicture}[scale=0.95]
        \begin{axis}
        [
        ybar,
        ytick={0,200,400,600,800,1000,1200},
        width=8cm,
    height=4cm,
         tick label style={/pgf/number format/1000 sep=},
         legend style={at={(0.05,0.95)},anchor=north west},
        ylabel={\# Repos},
        xlabel=Year,
        xlabel style={yshift=0pt},
        legend style={at={(0.5,-0.5)},anchor=north,legend columns=-1},
        nodes near coords,
        every node near coord/.append style={rotate=90, anchor=west, font=\scriptsize},
        bar width=5pt,
        width=\linewidth,
        grid style=dashed,
        ymajorgrids=true,
        ]
        
        \addplot[fill=archtGrey] coordinates {(2013, 11) (2014, 40) (2015, 67) (2016, 70) (2017, 45) (2018, 39) (2019, 33) (2020, 33) (2021, 51) (2022, 76) (2023, 193)};
      
        \addplot[fill=archtBlue]  coordinates {(2013, 61) (2014, 215) (2015, 310) (2016, 265) (2017, 177) (2018, 110) (2019, 124) (2020, 109) (2021, 163) (2022, 258) (2023, 546)};

        \addplot[fill=archtOrange] coordinates {(2013, 72) (2014, 249) (2015, 407) (2016, 388) (2017, 272) (2018, 199) (2019, 191) (2020, 217) (2021, 267) (2022, 526) (2023, 1093)};
        \legend{MPI+OpenMP, MPI, OpenMP}
        \end{axis}
        
        \end{tikzpicture}
        \caption{MPI, OpenMP, and MPI+OpenMP usage in \hpcorpus{} over the last decade.}
        \label{fig:omp-mpi}
\end{figure*}
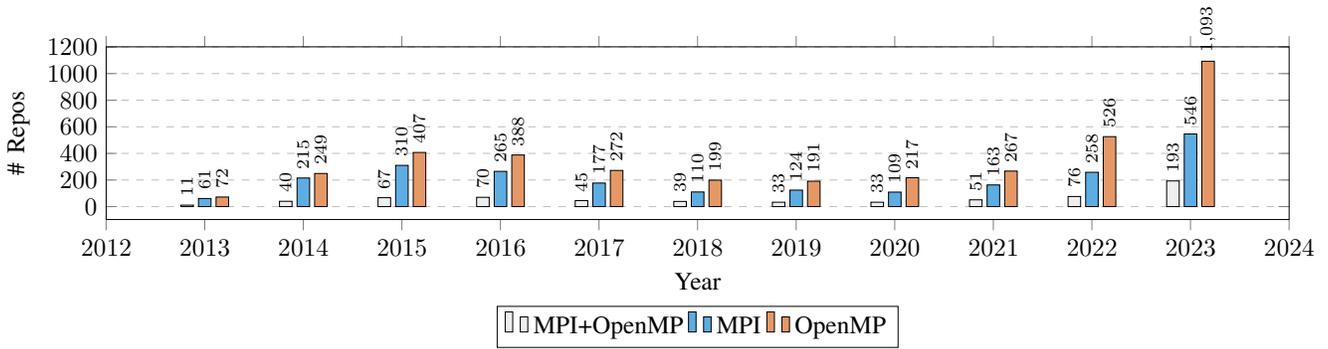
\subsection{Breakdown of OpenMP Directives Usage}

In this section, we discuss the detailed usage breakdown of elements from OpenMP as presented in \autoref{fig:fullbreakdown}\footnote{See code @ \href{https://github.com/Scientific-Computing-Lab-NRCN/HPCorpus/blob/121b0f088f3187cf82a54016e82992400cfcfe2c/generate_stats.py\#L275}{aggregate\_versions}.}.
This data breaks down usage statistics by language (C, C++, and Fortran) and by the version of the OpenMP specification (versions 2.5 to 5.2).\footnote{For complete OpenMP specification and the differences between versions, see \url{https://www.openmp.org/spec-html/5.0/openmp.html} and \cite{de2018ongoing}.}. The counts measure each clause's usage, with a higher count suggesting more prevalent usage in C, C++, and Fortran codes from \hpcorpus{}. 

Generally, over the years, we see that while new directives were well received and adopted (\autoref{fig:omp_versions}),\footnote{See code @ \href{https://github.com/Scientific-Computing-Lab-NRCN/HPCorpus/blob/121b0f088f3187cf82a54016e82992400cfcfe2c/generate_stats.py\#L204}{get\_version\_per\_year}.} the \textit{common core} of OpenMP v2.5-3, specifically OpenMP's \texttt{parallel for} constructs (\autoref{fig:fordoomp}),\footnote{See code @ \href{https://github.com/Scientific-Computing-Lab-NRCN/HPCorpus/blob/121b0f088f3187cf82a54016e82992400cfcfe2c/generate_stats.py\#L239}{get\_loops}.} dominate the usage. In addition to this analysis, we also analyze \hpcorpus{} for the growth in complexity  of OpenMP specifications (\autoref{fig:omp-specs-api}).\footnote{We refer the reader to a previous, well-known graph, which measured the growth in complexity by the specification page count (for example, reference \cite{mattson2019openmp}, Figure 3.1, page 37.)}  


 \input{directives}

\textbf{\texttt{for} loops:} \texttt{for} loops are commonly-used iterative control structures in programming. In C codes from \hpcorpus{}, there are 21,822,609 occurrences of for loops, followed by 19,841,237 in C++ codes and 1,203,773 (\texttt{do} loops) in Fortran. Fortran leads in parallelizing a proportion of \texttt{for} loops with OpenMP, with 2.2\% of all loops parallelized, compared to 0.54\% in C++ and 0.16\% in C.

\textbf{Scheduling:} Static scheduling is the preferred kind of schedule across all languages, with C codes from \hpcorpus{} having 8,705 occurrences, followed by C++ codes with 4,897 and Fortran codes with 2,013. Dynamic scheduling is less prevalent but still used more than advanced methods such as \texttt{guided}, \texttt{runtime}, and \texttt{auto}.

\textbf{Data-sharing:} C++ codes from \hpcorpus{} demonstrate higher usage of data-sharing constructs (\texttt{private}, \texttt{firstprivate}, \texttt{lastprivate}, \texttt{shared}, and \texttt{reduction}) compared to C and Fortran. However, the \texttt{nowait} construct, for eliminating synchronization barriers, is used less frequently in all languages.

\textbf{Irregular Parallelism:} In \hpcorpus{}, the \texttt{task} construct for explicit tasks and the \texttt{sections} construct for parallel sections are moderately prevalent. C++ codes from \hpcorpus{} have the highest counts with 4,169 \texttt{task} occurrences and 7,038 \texttt{sections} occurrences.

\textbf{Vectorization:} C++ codes from \hpcorpus{} have significantly higher usage of the \texttt{simd} directive for enabling vectorization, with 54,557 occurrences. In comparison, C and Fortran codes have 9,942 and 1,199 occurrences, respectively.

\textbf{Offloading:} The \texttt{target} construct for offloading computations to accelerators is prominently used in C++ codes in \hpcorpus{}, with 78,930 occurrences. C and Fortran codes, on the other hand, show 6,199 and 2,532 occurrences, resp.

\textbf{Synchronization:} The \texttt{barrier} construct for synchronization varies in usage, with C, C++, and Fortran codes from \hpcorpus{} having 2,825, 2,728, and 858 occurrences respectively. The \texttt{atomic} construct for atomic operations is moderately used, with C++, C, and Fortran codes, having 5,360, 3,601, and 2,005 occurrences, respectively. Other constructs such as \texttt{flush}, \texttt{single}, and \texttt{master} have relatively lower counts across all languages in \hpcorpus{}.




\newpage
\section{Discussion}
 
These results show that OpenMP is the dominant parallel programming model for C, C++, and Fortran in the publicly visible repositories in GitHub. The usage of items within OpenMP, however, is not evenly spread between the different versions of the OpenMP specification.   These results largely confirm the subset of 21 items from the OpenMP Common Core~\cite{mattson2019openmp} with one notable exception. The Common Core omitted the \texttt{lastprivate} clause. It was felt that this clause was rarely used.  Clearly, that is not the case.  Future updates of the Common Core should include it.    

While the Common Core is an important simplification of OpenMP, programmers often gravitate towards even greater simplicity.  The third most common construct used in OpenMP is \texttt{parallel for}.  This supports a style of programming where you find the time-consuming loops and then parallelize them with a simple \texttt{parallel for} directive. Much greater performance is available by explicitly managing parallel regions inside of which are worksharing loops (i.e., using the \texttt{for} construct).  The popularity of \texttt{parallel for} is a reminder to programming model designers; people often seek \textit{good-enough} performance, not ultimate performance.  

Finally, we wish to comment on the insights from the \hpcorpus{} data set on the adoption of new releases of OpenMP.   These new releases are not ignored.  New features of OpenMP find programmers that use them.  The fact of the matter is, however, that most OpenMP programmers work with items from OpenMP 4.0 or earlier (note: OpenMP 4.0 came out over 10 years ago).   The adoption of new features is uneven across the programming community.

\section{Conclusions and Future Outlook}

The central conclusion of this paper is that OpenMP is far more popular than we anticipated.  We expected the popularity of programming models in \hpcorpus{} to mirror that found in workloads at various supercomputer centers.   For example, a 2015 talk at OpenMPcon~\cite{commonstats} reported that over 90\% of applications programs running at NERSC used MPI, while only 40\% used OpenMP.  

We found, however, that for code in GitHub, OpenMP is by far the most popular parallel programming model for C, C++, and Fortran, with 45\% of repositories containing OpenMP code while only 27\% containing MPI.  This result is so surprising that we can't help but wonder if we did something wrong. We have made \hpcorpus{} available and shared the query used to generate it.  We are eager for peer review to help ensure that our conclusions are correct.

In retrospect, however, perhaps we should not have been so surprised.  Programs running at supercomputer centers  are designed to run on large scalable machines.  These programs obviously need a distributed memory API such as MPI.  On GitHub, however, you would expect to find distributed memory programs but also multithreaded programs for a wide range of multicore CPUs.  In other words, the scope for OpenMP extends from edge devices to laptops, to servers, to massive supercomputers.  The number of systems that benefit from OpenMP dwarfs the number of large-scale clusters or even GPU-based systems; hence it is not surprising that the data from \hpcorpus{} shows so much usage of OpenMP.

When you work with an MPI program, you know it.  The program is launched as an MPI program (with \texttt{mpiexec} or \texttt{runmpi}).  The code is, in most cases, structured around a single program multiple data (SPMD) pattern, with a copy of the program running on each node of the system.  You know you are running an MPI job.  With OpenMP, it is easy to use it and not even know it.  A library routine buried deep in your code could use OpenMP.  It's much lower overhead to experiment with OpenMP compared to MPI.  Hence, the use of MPI stands out, and frankly, it is easy to forget about OpenMP.  This factor could also play a role in expectations of MPI usage relative to OpenMP.

There is much work yet to do with \hpcorpus{}. To drive the development of new  programming models, we need to understand how programming models are used.  To do this we need to move beyond counts for the different
constructs and explore the different ways they are combined into distinct design patterns of parallel 
programming~\cite{mattson2004patterns}.  From these patterns, we can better understand the cognitive issues of programmers working with different parallel programming models and guide how they should evolve.

Our interest in \hpcorpus{} goes well beyond studying which constructs are used in different parallel programming 
models. \hpcorpus{} can be used to train LLMs to support AI solutions for
machine programming.  Even with the amazing results from generative AI applied to programming, we
are in the early days of this technology.  We are interested in training LLMs to address different programming problems.   While the ``black box'' that is current AI technology is fascinating,
we see the line of research growing out of \hpcorpus{} going much further into models that combine neural networks with symbolic systems to reason about correctness while working with high-level structures based on fundamental design patterns of parallel programming.  Only by combining neural networks with symbolic reasoning will we be able to crack the machine programming problem.

In closing, we call on others to carry out studies similar to the one described in this paper.   For example, other than the talk about workloads at NERSC~\cite{commonstats}, we were unable to find detailed studies of programming model usage at major supercomputing centers.  We focused on C, C++, and Fortran, but it would be interesting to repeat this work for Python, Rust, and Julia.  Code repositories have become the standard way to manage complex software projects.  The data is ``out there''.   We should learn what programmers are actually using and then drive the evolution of programming models based on hard data, not anecdotes. 

\clearpage
\section*{Acknowledgments}
This research was supported by the Israeli Council for Higher Education (CHE) via the Data Science Research Center, Ben-Gurion University of the Negev, Israel; Pazy grant 226/20; Intel Corporation (oneAPI CoE program); and the Lynn and William Frankel Center for Computer Science. Computational support was provided by the NegevHPC project~\cite{negevhpc}, Intel Developer Cloud~\cite{intel-cloud}, and Google Cloud Platform (GCP). The authors thank Re'em Harel, Yehonatan Fridman, Israel Hen, and Gabi Dadush for their help and support.

\bibliographystyle{IEEEtran}
\bibliography{IEEEabrv,sample-base.bib}

\end{document}